# Planar Atom Trap and Magnetic Resonance 'Lens' Designs

Mladen Barbic* and Christopher P. Barrett
*Department of Physics and Astronomy, California State Univ. Long Beach, 1250 Bellflower Boulevard, Long Beach, CA 90840, USA*
and
Teresa H. Emery and Axel Scherer
*Department of Electrical Engineering, California Institute of Technology 1200 East California Boulevard, Pasadena, CA 91125, USA*

**Abstract**

We present various planar magnetic designs that create points above the plane where the magnitude of the static magnetic field is a local minimum. Structures with these properties are of interest in the disciplines of neutral atom confinement, magnetic levitation, and magnetic resonance imaging. Each planar permanent magnet design is accompanied by the equivalent planar single non-crossing conductor design. Presented designs fall into three categories producing: a) zero value magnetic field magnitude point minima, b) non-zero magnetic field magnitude point minima requiring external bias magnetic field, and c) self-biased non-zero magnetic field magnitude point minima. We also introduce the Principle of Amperean Current Doubling in planar perpendicularly magnetized thin films that can be used to improve the performance of each permanent magnet design we present. Single conductor current-carrying designs are suitable for single layer lithographic fabrication, as we experimentally demonstrate. Finally, we present the case that nanometer scale recording of perpendicular anisotropy thin magnetic films using presently available data storage technology can provide the ultimate miniaturization of the presented designs.

**Introduction**

It has been well recognized that Maxwell's equations place interesting restrictions on the properties of magneto-static fields in free space. For example, the magnitudes of individual components of the magnetic field vector cannot have a local minimum or maximum in free space. Additionally, the magnitude of the magnetic field, $|B|$, cannot have a local maximum, but can interestingly have a local minimum in free space and that minimum can have a non-zero value. These properties and their consequences have been studied for a long time [1,2] and have received much attention in numerous scientific disciplines such as plasma confinement [3,4], neutral particle trapping [5-12], diamagnetic levitation [13-20], and magnetic resonance imaging [21,22]. Of particular interest to us are microscopic planar structures in the form of non-crossing current-carrying conductors or thin film permanent magnet materials. As such, they would achieve the ultimate miniaturization of the magnetic field magnitude minima regions in free space. In turn, this would provide the tightest confinement potentials for neutral atoms in atom chip applications [9], as well as create the smallest focus region of a magnetic resonance 'lens' for potentially atomic resolution imaging performance [22].

We present both planar permanent magnet and planar current carrying conductor designs. Permanent magnet designs have the advantage that, when miniaturized, they produce higher fields than an electromagnet and can be scaled down to any size without any loss in field strength [23]. Miniaturized permanent magnets also provide larger magnetic field gradients and curvatures than their current-carrying conductor counterparts, require no outside power supply and no interconnecting leads, and finally, generate no heat and require no heat dissipation. We utilize Ampere's Principle of Equivalence between the magnetization distribution and current loops [24] in constructing current carrying structures equivalent to their magnetic counterparts. Current-carrying conductor designs might be useful when tuning of the magnetic fields is desired. We restrict our current carrying structures to planar single non-crossing conductor designs in order to minimize the complexity of electrical connections to the structure, as well as make them suitable for standard single layer lithographic fabrication. We restrict our permanent magnet designs to planar perpendicularly magnetized thin films in order to make our structures realizable in presently available perpendicular anisotropy magnetic thin films.

We model the magnetic fields above thin perpendicular anisotropy magnetic films of 10nm thickness and uniform magnetization **M** of $\mu_0 M = 2$(Tesla) directed along the z-axis. Therefore, the uniform positive magnetic pseudo-surface-charge density of **n·M** is on the top surface of the thin film, where **n** is the outward normal of the magnet, and the corresponding negative pseudo-surface-charge density is on the bottom surface of the thin film. We numerically compute first the scalar potential $\varphi(\mathbf{r})$ and then the magnetic field **B(r)** at a position **r** above the plane:

$$\vec{B}(\vec{r}) = -\mu_0 \nabla \varphi(\vec{r}) = -\mu_0 \nabla \left( \sum_{i=1}^{2} \frac{1}{4\pi} \int_{S_i} \frac{\hat{n}_i \cdot \vec{M}_i}{|\vec{r} - \vec{r}\,'|} da' \right) \qquad (1)$$

where $\mu_0$ is the permeability of free space. All of our presented designs can in principle be recorded in the perpendicular thin films by utilizing the presently available data storage technology. The idea of recording the described designs has the important advantages of using the Principle of Amperean Current Doubling that we will describe for improving the magnetic performance of the structures in infinite thin magnetic films, while altogether avoiding lithographic fabrication.

**Zero Magnitude Minima Structures**

In Figure 1 we show the simplest examples of planar magnetic structures that provide out of plane local magnetic field magnitude minima of value zero. The principles of these designs are not novel, as they have been previously described in both current carrying conductor [9] and permanent magnet forms [25-26]. We show them for completeness, as well as to describe single non-crossing planar conductor designs that might be useful in practical situations where the simplicity of single layer lithographic fabrication is desired. A simple thin permanent magnet circular disk magnetized perpendicular to the plane of the disk (out of the page) is shown in Figure 1a, with the arrows showing the equivalent Amperean pseudo-currents. This structure requires a uniform external bias magnetic field oriented opposite to the magnetization of the disk in order to produce an out of plane magnetic field magnitude minimum of zero value. Figure 1b shows the single conductor current-carrying design that is equivalent to the permanent magnet structure of Figure 1a.

A self-biased zero value local magnetic field minimum permanent magnet structure is shown in Figure 1c. The magnetic material is again magnetized perpendicular to the plane, and the Amperean pseudo-currents of the structure are indicated by the arrows. Figure 1d shows the single non-crossing planar current carrying conductor equivalent of the structure shown in Figure 1c. The magnetic fields from adjacent straight wire segments of the device are approximately cancelled in the out of plane minima region of interest, as the currents in those straight sections flow in the opposite directions. The single non-crossing conductor design we show simplifies the complexity of the device, but has the limitation of being able to apply only equal current throughout the entire device. The optimum arrangement would be a structure with two unequal currents flowing in the opposite directions [9], but such a design requires multiple conductors, multiple current sources, and multiple lithographic fabrication steps. Simple analysis shows that our single non-crossing conductor design is optimized for the maximum trap depth when the outer diameter is 2.2 times the inner diameter, as in the structure shown in Figure 1d.

### Biased Non-Zero Magnitude Minima Structures

Designs that produce local non-zero magnetic field magnitude minima are of special interest as they provide improved performance in neutral atom trapping [6,9], and provide the opportunity for local magnetic resonance spectroscopy and imaging [21,22,27,28]. We present a couple of simple designs, while noting that our example structures are chosen for their simplicity and the ability to obtain single planar non-crossing conductor analogues that achieve the desired magnetic field properties. Our designs are by no means exhaustive, and as suggested by Wing [5]: "*No doubt ingenuity will reveal a number of other interesting geometries as well.*"

Figure 2a shows perhaps the simplest permanent magnet design with the Amperean pseudo-currents indicated. For example, for inner radius being 40 nanometers, outer radius 60 nanometers, the structure creates a local magnetic field magnitude minimum centered on the z-axis, and 55nm above the plane with a magnetic field magnitude minimum value of 19 Gauss. In order to create the local magnetic field magnitude minimum above the plane, a uniform external bias magnetic field of 100 Gauss in direction opposite to the magnetization orientation is required. The single non-crossing conductor equivalent design is shown in Figure 2b. Again, the magnetic fields from adjacent straight wire segments of the device are approximately cancelled in the out of plane minima region of interest, as the currents in those straight sections flow in the opposite directions. As in the permanent magnet case, a uniform external bias field is required to generate a local magnetic field magnitude minimum.

Figure 2c shows another permanent magnet design, initially introduced as an MRI "Lens" [22]. Briefly, a local non-zero magnetic field magnitude minimum point above the structure can be used as a 'focus' spot where potentially only a single spin would be magnetically resonant and therefore detected. The structure creates a local magnetic field magnitude minimum with a value of 99.5 Gauss, with a required external uniform bias field of 650 Gauss in direction opposite to the magnetization direction. The location of the minimum is at z=24nm above the surface. In addition to the permanent magnet case, we show in Figure 2d the equivalent planar non-crossing single conductor design. It achieves the same magnetic field patterns in free space as long as the required uniform external bias magnetic field is applied.

In Figure 3 we show two examples of the nanofabricated structures (designs of Figure 1d and Figure 2d) using standard single layer electron beam nanolithography. The fabrication procedure included: a) spin coating of single layer electron beam resist on top of an insulator on silicon substrate, b) beam exposure in the Leica EBPG 5000+ electron beam lithography system, c) thermal evaporation of 10nm Chromium and 30nm of Gold, and d) lift-off revealing the final non-crossing single 100nm line-width conductor structures shown in Figure 3.

**Self-Biased Non-Zero Magnitude Minima Structures**

Structures that do not require an external magnetic field and produce local non-zero magnetic field magnitude minima above the plane would be of interest in both neutral atom trapping and magnetic resonance spectroscopy and imaging applications. Such structures would potentially benefit from the simplification of the experimental arrangement by removing the uniform external magnetic field, as well as from avoiding the potentially detrimental effects and interactions of that external field with other components of the experiment. We present the two simplest permanent magnet structures that we found for which there are planar single non-crossing conductor equivalents.

Figure 4a is a modification of the design of Figure 2c where the outer effective Amperian pseudo-current provides the required bias field, although not a uniform one. With the outermost radius of 150nm, the structure has a local magnetic field magnitude minimum with value of 75 Gauss centered on z-axis at z=28nm. The stated dimensions give the highest value of the magnetic field magnitude minimum. Smaller outermost diameter values result in the creation of saddle points instead of localized true minima. This structure has a planar single non-crossing conductor equivalent, shown in Figure 4b.

The second self-biased structure with better performance with respect to the magnetic field magnitude minimum value and its location above the plane, as well as the amount of magnetic material required is shown in Figure 5a. For the innermost radius of 40nm, middle radius of 60nm, and outermost radius of 220nm, the structure has a local magnetic field magnitude minimum of 120 Gauss at z=66.5nm. This structure also has a planar single non-crossing conductor equivalent shown in Figure 5b. In Figure 6 we again demonstrate the single layer electron beam nanofabrication of the proposed designs of Figure 4b and Figure 5b, respectively.

**Principle of Amperean Current Doubling**

Inspection of Equation 1 reveals that the choices of magnetic materials used in the construction of the presented designs as well as scaling by nanofabrication could be utilized to enhance the performance of the devices. By enhancement, we generally mean the increase of the values of the magnetic field magnitudes, magnetic field gradients, and magnetic field curvatures in the region of the local magnetic field minima above the plane of the structure. Such enhancements would provide tighter confining potentials in neutral atom trapping applications [6,9], and higher spin polarization, higher nuclear magnetic resonance precession frequency, and higher spatial resolution in magnetic resonance spectroscopy and imaging [22]. While it is difficult to achieve saturation magnetization of thin perpendicular anisotropy magnetic films above our simulation value of 2 Tesla, we present here a technique we term the Principle of Amperean Current Doubling that can be used to double the magnetic field parameters of interest in structures recorded in infinite thin magnetic films.

The Principle of Amperean Current Doubling is described through Figure 7. As a representative design, we show the structure of Figure 2c, although we emphasize that the present argument is valid for all the structures presented in this article. In Figure 7a the thin permanent magnet film is magnetized perpendicular to the plane of the film (*out of the page*) and the equivalent Amperean pseudo-currents of the magnetic structure are shown as arrows on the boundaries of the structure. As a reminder, a uniform external bias field in direction opposite to the magnetization direction of the thin film (*into the page*) is required to produce a magnetic field magnitude minimum above the plane of the film. In Figure 7b we show an infinite magnetic thin film structure that has the same equivalent Amperean pseudo-currents as the structure of Figure 7a and therefore produces the same magnetic fields, the same magnetic field gradients, and the same magnetic field curvatures as the structure of Figure 7a. In the case of this second structure, the infinite thin film with perpendicular magnetic anisotropy is magnetized *into the page*. Interestingly, this structure requires the same uniform external bias magnetic field in order to produce the magnetic field magnitude minimum out of the plane of the film, and therefore the uniform external bias magnetic field is along the same direction as the magnetization direction of the thin film (*into the page*).

The structures of Figure 7a and Figure 7b are therefore exactly equivalent in terms of the magnetic fields they produce with the interesting caveat that they are the *exact spatial negatives* of one another. This is important, as it means that they can be added to form a single composite structure, shown in Figure 7c, in which their effective Amperean pseudo-currents add so that the total effective current is doubled, as indicated by the larger arrows on the magnetic transitions of the infinite thin film structure. Therefore the values of the magnetic fields, magnetic field gradients, and magnetic curvatures of the structure shown in Figure 7c are doubled as compared to the structure shown in Figure 2c. The location of the local magnetic field minimum remains at the same location of $z=24$nm, but the value of the field magnitude at the local minimum is doubled to 199 Gauss, and the gradients and field curvatures are also doubled. It is important to mention that the required external uniform bias field also has to be doubled to 1,300 Gauss in order to create this local non-zero minimum.

This Principle of Amperean Current Doubling might have important consequences, for example, in the use of this structure as a potential MRI 'Lens' [22] where the doubling of the field will provide the doubling of the nuclear spin polarization as well as the doubling of the nuclear spin magnetic resonance precession frequency, both of which are advantageous and would in principle result in quadrupling of the signal to noise ratio in inductive detection of nuclear magnetic resonance. The effect of current doubling on the magnetic field gradients and curvatures at the local non-zero minimum would also result in improvement of the spatial imaging resolution of such a 'Lens', as well as in the increase of forces on resonant spins for the case of force detected magnetic resonance [22].

**Recorded Square Form Factor Designs**

Although magneto-optical data storage technology has already been utilized for recording of simple atom trapping structures with zero value local minima [25,26] that we showed in Figure 1, we here explore the ultimate miniaturization possibilities with respect to all of the structures we described. We present the case that recording of perpendicular anisotropy thin magnetic films using presently available data storage technology can provide the ultimate miniaturization of our presented designs. We believe that with such technology it is possible to: a) experimentally implement the idea of Amperean Current Doubling, b) achieve nanometer precision recording of the described designs in a square form factor, c) record the presented designs with nanometer resolution over large (centimeter scale) areas in a reasonable time, and d) avoid altogether potentially expensive lithographic fabrication of the presented designs.

Presently available information data storage technology has already been successfully utilized in writing of various magnetic patterns of relevance to magnetic recording science using Scanning Magneto-Resistance Microscopy [29-31]. The attractive feature of this technique is that magnetic patterns of nanometer positioning precision can be recorded with an inductive write element, and then read-back with a magneto-resistive read element in a raster scanning fashion in order to produce an image of the recorded pattern [32-38]. The technique has been extended to high-speed recording and imaging of patterns in thin magnetic film media of both longitudinal and perpendicular form over large-scale areas [39-45] using a well-established magnetic recording test and evaluation instrument called 'Spinstand' [46].

By recording our presented designs in infinite perpendicular thin film magnetic media using such techniques, the desirable goals of miniaturization and Amperean Current Doubling would be accomplished on optically smooth large-scale areas while altogether avoiding lithographic nanofabrication. The only limitation of these recording methods is that the magnetic patterns are of a square form factor, although this is in no way prohibitive with respect to the arguments presented in this article. For example, our numerical modeling shows that recording of the patterns shown in Figure 8a and 8b (square form factor versions of circular designs shown in Figure 4a and 5a, respectively) would result in successful creation of out-of-plane self-biased non-zero value local magnetic field magnitude minima of similar parameter values as achieved with circular designs. Such patterns of sub-micron overall dimensions could be recorded with nanometer scale precision (if needed over large spatial areas) based on the experimentally demonstrated published reports [32-45] and specifications of the state-of-the-art Spinstand products [46].

## Conclusion

In summary, we have presented various perpendicular anisotropy magnetic thin film and planar non-crossing current carrying conductor designs that create local point minima of the magnetic field magnitudes above the plane of the structures. We demonstrated electron beam defined nanofabrication of some of the planar conductor designs. With further advances in nanolithography, thin film magnetism, and magnetic recording and sensing technologies, these proposed structures may provide the ultimate confinement potentials and "focusing" regions desired in atom trapping and magnetic resonance microscopy applications.

## Acknowlegment

This material is based upon work supported by the National Science Foundation under the NSF-CAREER Award Grant No. 0349319 and NSF-ECS Award Grant No. 0622228, as well as the California State University Long Beach Scholarly and Creative Activities Award (SCAC).

**Figure Captions**

Fig. 1. a) A thin magnetic circular disk magnetized perpendicular to the plane and a uniform external bias magnetic field oriented opposite to the magnetization produce an out of plane magnetic field magnitude minimum of zero value. The arrows show the equivalent Amperean pseudo-currents. b) Single conductor current-carrying design that is equivalent to the permanent magnet structure of a). c) A self-biased zero value local magnetic field minimum permanent magnet structure. The magnetic material is magnetized perpendicular to the plane, with the Amperean pseudo-currents of the structure indicated by the arrows. d) Single non-crossing planar current carrying conductor equivalent of the structure shown in c).

Fig. 2. a) A thin permanent magnet structure that produces an out of plane magnetic field magnitude point minimum of non-zero value. External uniform bias field opposite to the magnetization direction is required. The arrows show the equivalent Amperean pseudo-currents. b) Single conductor current-carrying design that is equivalent to the permanent magnet structure of a). c) Another non-zero value local magnetic field minimum permanent magnet structure producing higher value of the magnitude point minimum. The structure also requires a uniform external bias magnetic field. The Amperean pseudo-currents of the structure are indicated by the arrows. d) Single non-crossing planar current carrying conductor equivalent of the structure shown in c).

Fig. 3. Two examples of the fabricated 100nm line width single planar non-crossing conductor structures (designs from Figure 1d and Figure 2d) using standard single layer electron beam nanolithography.

Fig. 4. a) Self-biased perpendicular anisotropy magnetic thin film design producing local non-zero magnetic field magnitude point minimum above the plane of the structure. b) Planar single non-crossing conductor equivalent of the structure shown in a).

Fig. 5. a) The second self-biased perpendicular anisotropy magnetic thin film design producing local non-zero magnetic field magnitude point minimum above the plane of the structure. This structure has a larger magnetic field magnitude minimum value and its location above the plane is higher than the structure shown in Figure 4a. b) Planar single non-crossing conductor equivalent of the structure shown in a).

Fig. 6. Examples of the fabricated 100nm line width single planar non-crossing conductor structures (designs from Figure 4b and Figure 5b) using standard single layer electron beam nanolithography.

Fig. 7. The principle of Amperean current doubling in thin perpendicular anisotropy magnetic film structures. a) The design of Figure 2c with magnetization out of the page and its equivalent Amperean pseudo-currents indicated with arrows. b) Infinite thin film structure magnetized into the page has the same equivalent Amperean pseudo-currents as the structure shown in a) and therefore produces the same magnetic fields. Being exact spatial negatives of one another, the structures of a) and b) can be added to form a single composite structure, shown in Figure 7c, in which the total effective Amperean pseudo-currents are doubled.

Fig. 8. Square form factor version of the circular designs of Figures 4a and 5a, respectively, that can be achieved through recording of infinite perpendicular anisotropy thin magnetic films using data storage technology. For example, the black regions are recorded magnetized into the page, while the rest of the infinite thin film is magnetized out of the page. This would provide for Amperean current doubling and nanometer precision in recording on optically smooth large-scale areas, while avoiding lithographic fabrication. These self-biased square form factor structures produce out-of-plane non-zero value local magnetic field magnitude point minima of similar parameter values as achieved with circular designs of Figure 4a and 5a.

Figure 1

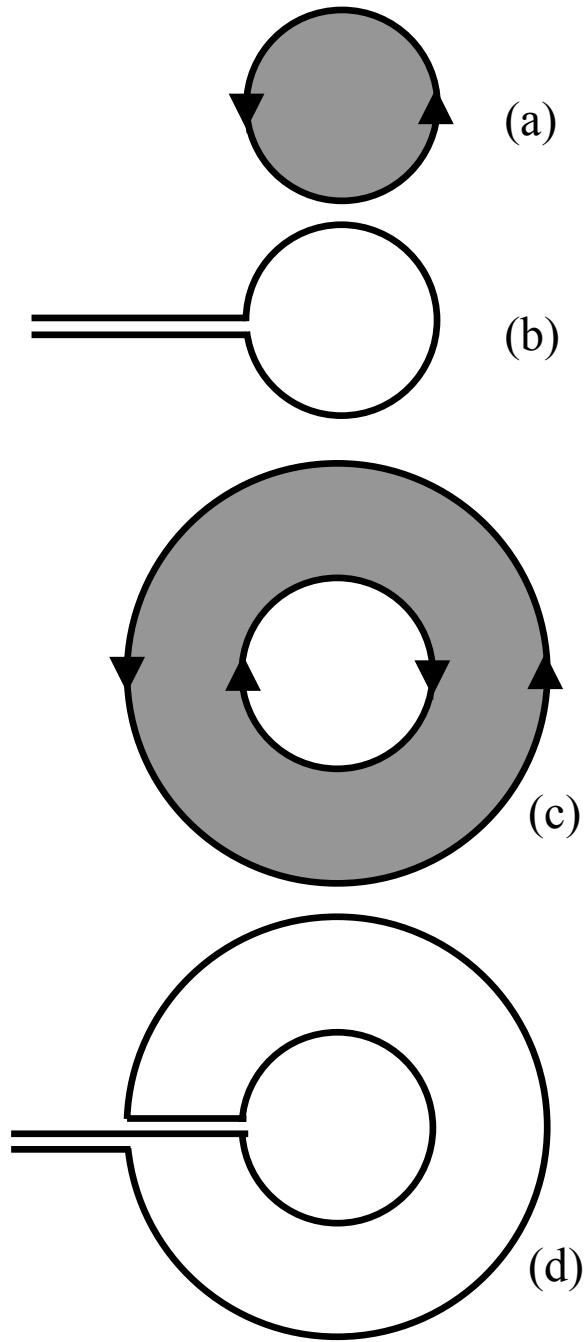

Figure 2

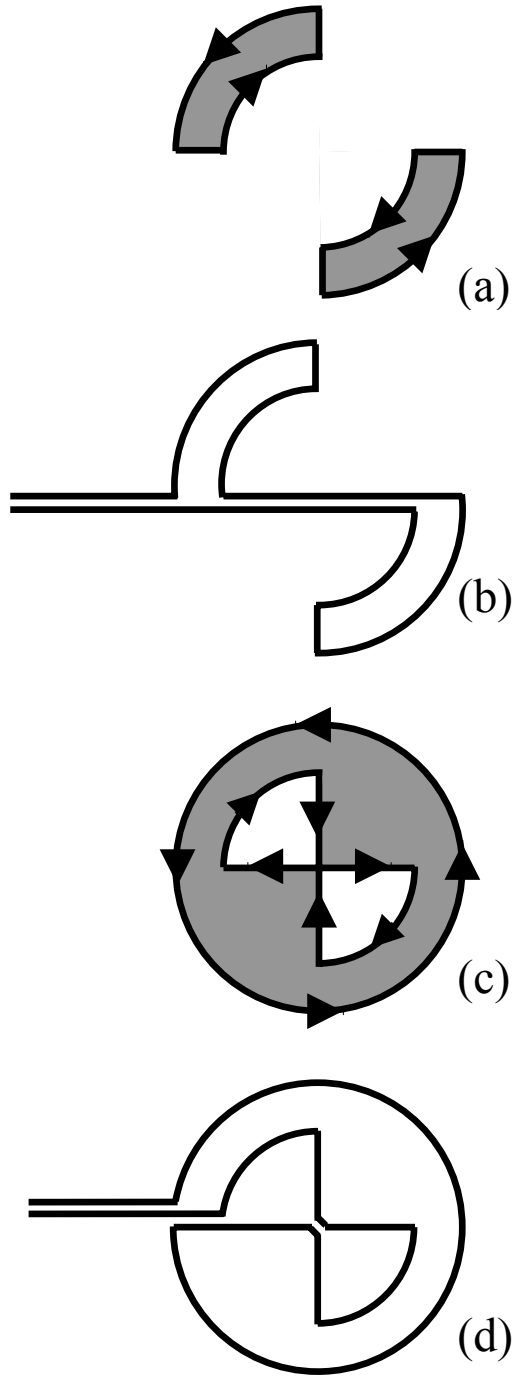

Figure 3

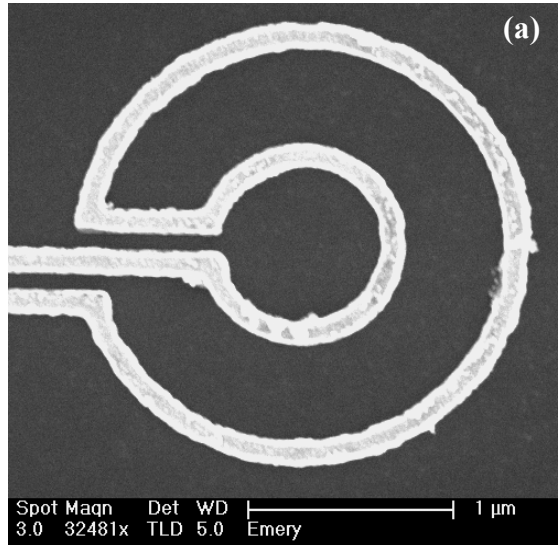

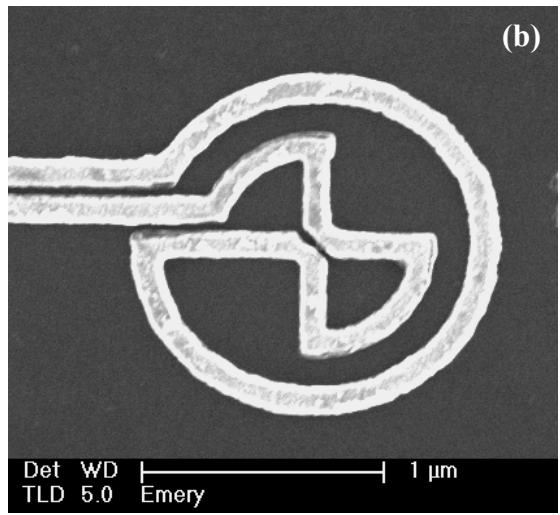

Figure 4

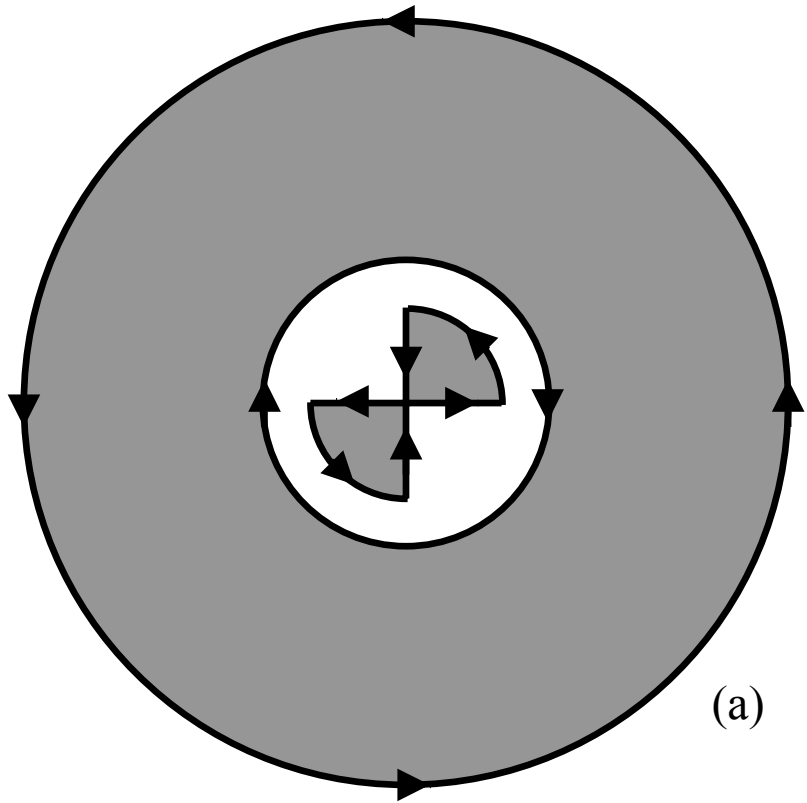

(a)

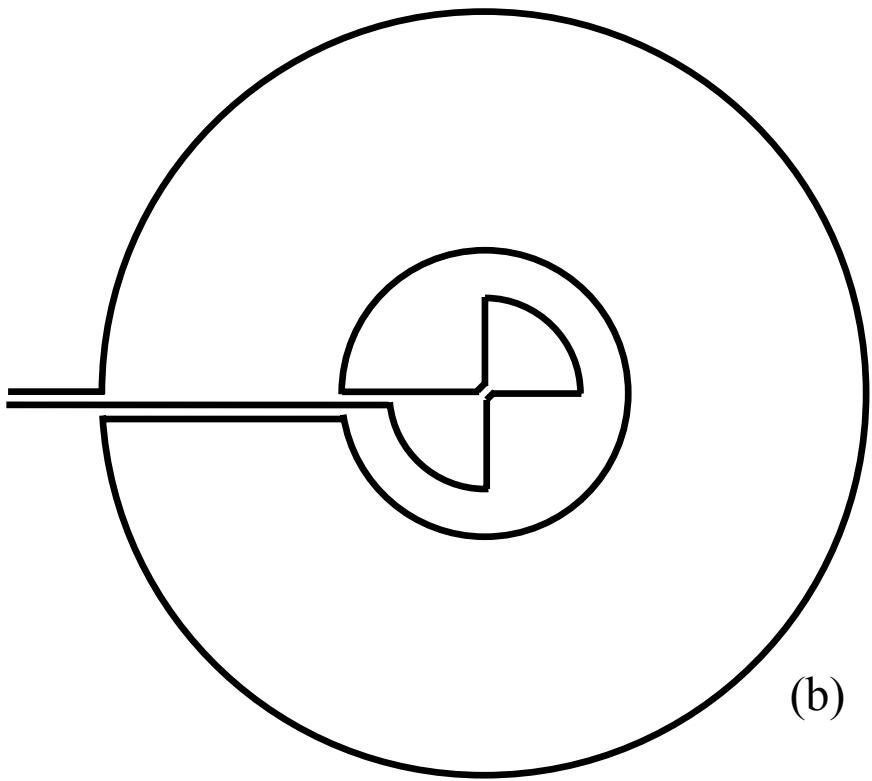

(b)

Figure 5

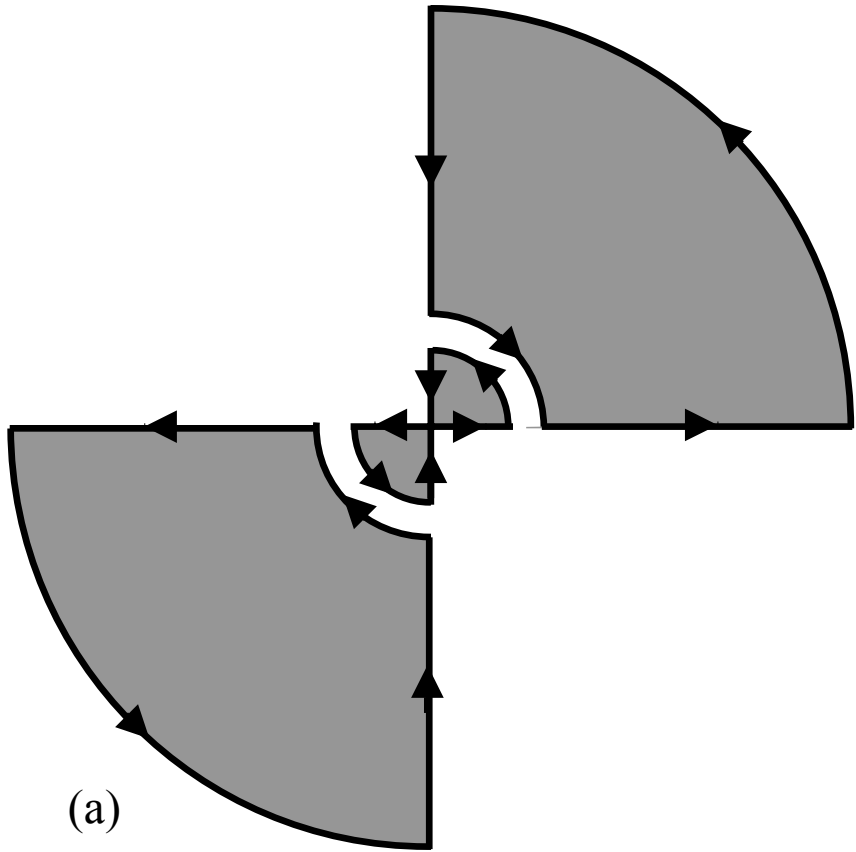

(a)

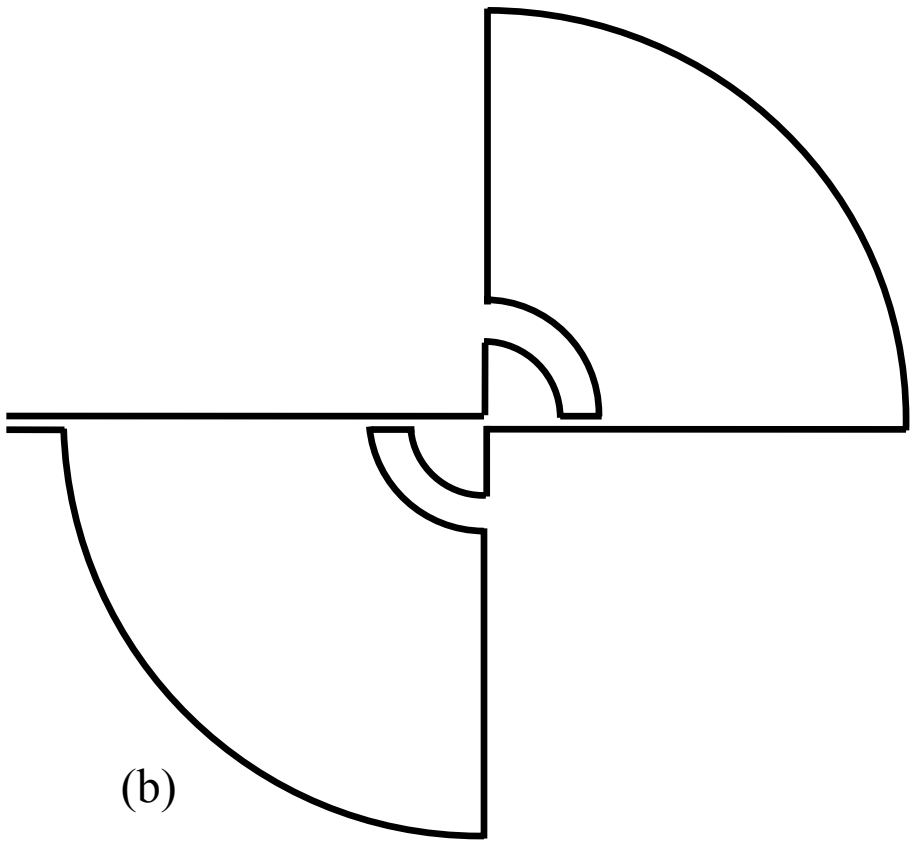

(b)

Figure 6

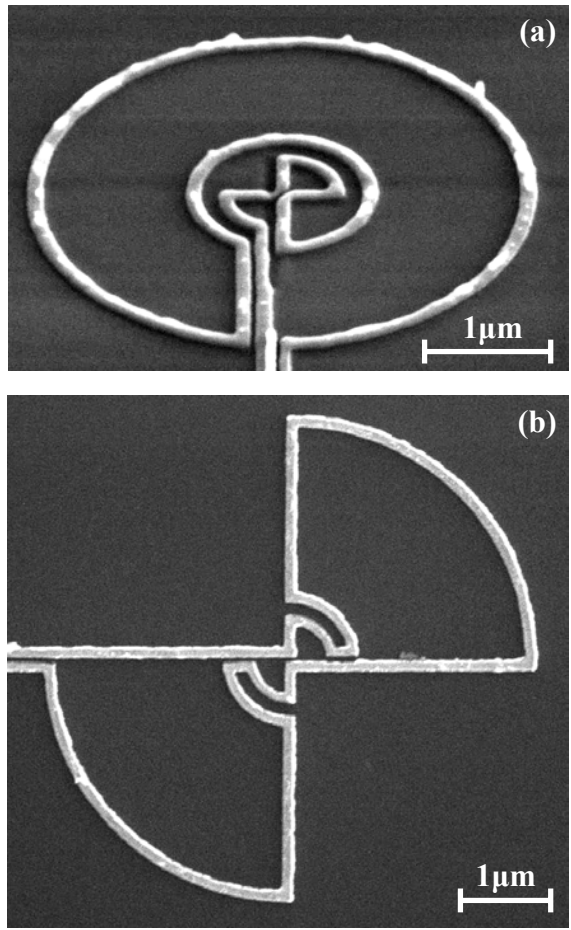

Figure 7

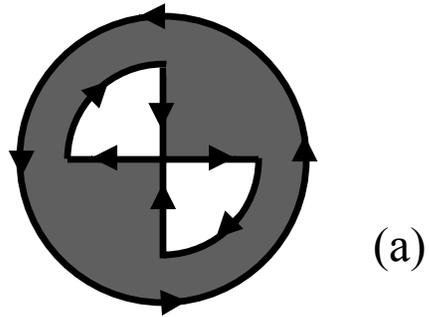

(a)

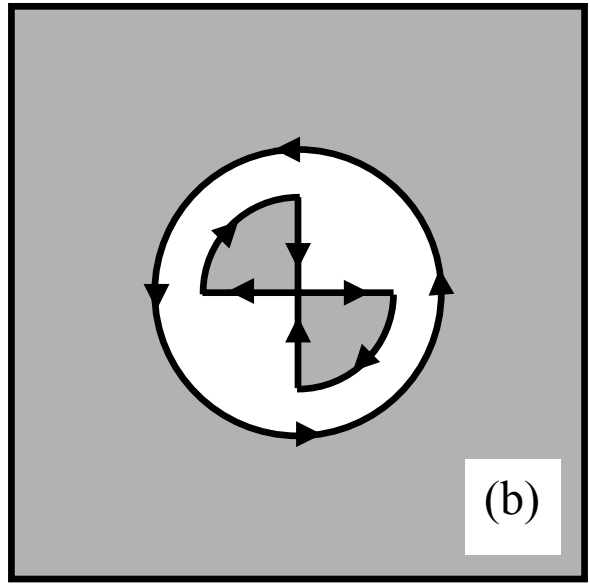

(b)

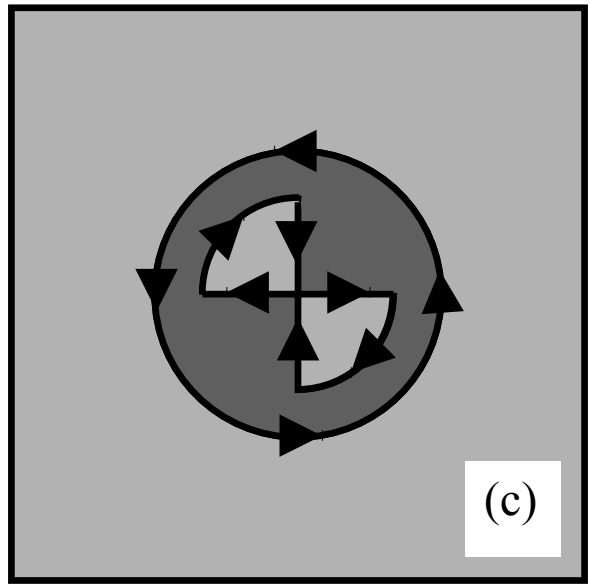

(c)

Figure 8

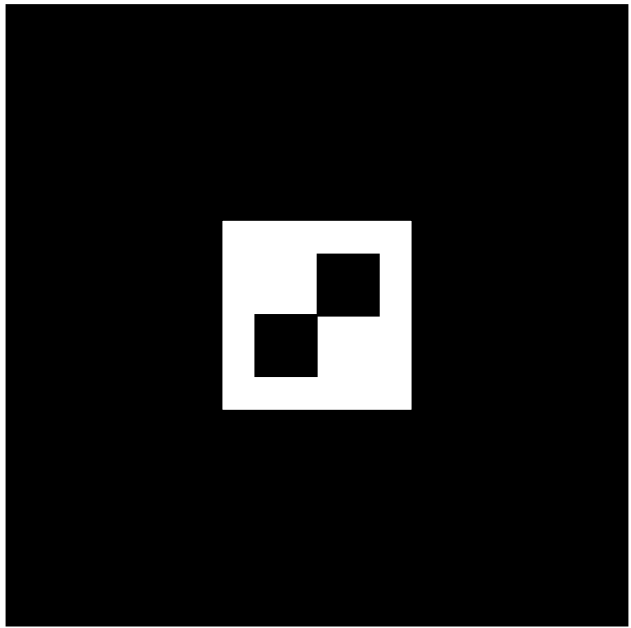

(a)

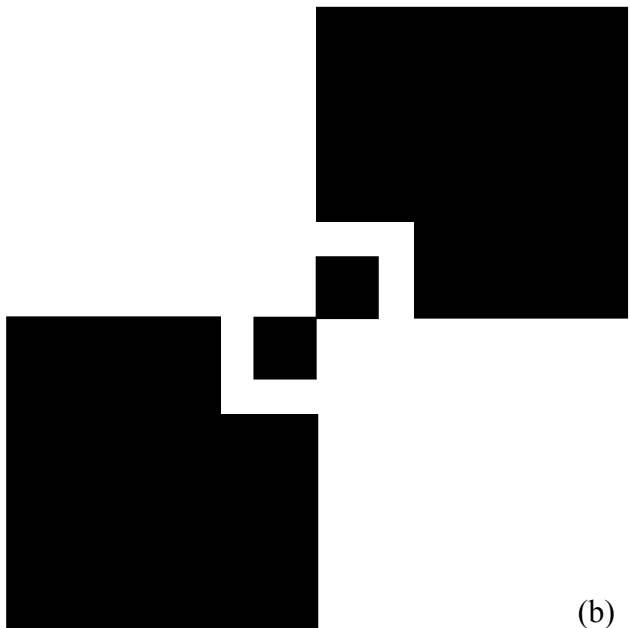

(b)